\documentclass[letter,twocolumn]{jpsj2} 
\usepackage{latexsym}
\usepackage{bm}
\usepackage{graphicx}

\title{
Topological meaning of Z$_2$ numbers in time reversal invariant systems
}

\author{Takahiro Fukui$^1$, Takanori Fujiwara$^1$, and Yasuhiro Hatsugai$^2$}
\inst{
$^1$Department of Physics, Ibaraki University, Mito
310-8512, Japan \\
$^2$
Institute of Physics, 
University of Tsukuba, 1-1-1 Tennodai, Tsukuba, Ibaraki 305-8571, Japan
}

\recdate{\today}

\abst{We show that the Z$_2$ invariant, which classifies the topological properties of 
time reversal invariant insulators, has deep relationship with the global anomaly.
Although the second Chern number is the basic topological invariant characterizing 
time reversal systems, we show that the relative phase between the Kramers doublet 
reduces the topological quantum number Z to Z$_2$.
}

\kword{Quantum spin Hall effect, Time reversal symmetry, 
Z$_2$ topological invariant, global anomaly}

\begin{document}
\maketitle

As one of possibilities for spintronics, spin Hall effect (SHE) has been proposed 
\cite{MNZ03,Sin04}, 
and much effort has been made to realize it experimentally \cite{KMGA04,WKSJ04,KWBetal07}.
In particular, the quantum spin Hall effect (QSHE) has been attracting much current 
interest due to its unique topological characteristics\cite{KanMel05a,KanMel05b,BerZha06}.
Although it may be regarded as a spin version of integer quantum Hall effect
(IQHE), their differences with respect to topological properties have been revealed.
It is well known that IQHE states are classified by integers which are 
topological invariant called first Chern number \cite{TKNN82,Koh85,HatEdge93,Fukui05}.
Kane and Mele \cite{KanMel05a} proposed that
time reversal invariant states are labeled by Z$_2$ numbers, and
much work has been done to clarify 
its meaning.\cite{XuMoo06,Roy06,FuKan06,Fukui07a,Fukui07b,MooBal07,RMOF07,LeeRyu08}
The Z$_2$ number implies that time reversal invariant systems have two phases:
One is the ordinary insulator, whereas the other is the topological insulator
\cite{FKM07,FuKan07} in which QSHE is expected.

On the other hand, two decades ago,  
Avron {\it et. al.}  \cite{ASSS88} argued that the basic topological invariant for
time reversal invariant systems is the second Chern number.
Qi, Hughes and Zhang 
recently proposed a generic topological field theory
in 4+1 dimensions, where basic topological properties are 
classified by the second Chern number\cite{QHZ08, HatFuk08}.
From this, they have constructed topological insulators
in two  and three spatial dimensions by the use of the dimensional reduction.

In this paper, we address the question what is the fundamental relationship between 
the second Chern number and Z$_2$ number. 
We show that time reversal invariant systems are indeed labeled by
the second Chern numbers.
However, taking account of the relative phase ambiguity
between a Kramers doublet, the topological classification by Z falls into 
two sectors labeled by Z$_2$. This reflects $\pi_4$(Sp($1$))=Z$_2$ 
(more generically, $\pi_4$(Sp($n$))=Z$_2$), and 
it has close relationship with the global anomalies found by Witten.
\cite{Witten83a,Witten83b} 

In generic Hamiltonian with broken time reversal symmetry, the eigenvalue 
degeneracies have codimension three.\cite{ASSS88} 
This means that due to level crossings, 
Hamiltonian depending on three parameters
could have some isolated ``monopoles'' in the parameter space, and 
on some two dimensional subspaces (typically the 2-sphere, $S^2$), 
the first Chern number takes a nontrivial integer if 
the monopoles exist inside the sphere.
On the other hand, in the time reversal invariant systems with ${\cal T}^2=-1$,
the eigenvalue degeneracies have codimension five.\cite{ASSS88,HatFuk08}
Therefore, to reveal the topological properties of such 
systems, we need to examine Hamiltonians with five parameters.
The minimal model may be the quaternionic 2$\times$2 hermitian matrix
or equivalently, 4$\times$4 hermitian matrix with time reversal invariance, 
which indeed depends on five parameters.\cite{ASSS88,HatFuk08}

Let $H(x)$ be the Hamiltonian satisfying ${\cal T}H(x){\cal T}^{-1}=H(x)$,
where $x$ denotes a set of five parameters $x_a$ ($a=1,\cdots,5$).
Each eigenstate is doubly-degenerate, which is referred to as Kramers doublet.
Such an eigenstate with the minimum twofold degeneracy may be called 
quaternionically simple.\cite{ASSS88} 
At certain points in five dimensional (5D)
parameter space, level crossings among several Kramers doublets occur. 
Then, we can consider a specific 4D subspace in which 
all the eigenstates are quaternionically simple, that is, 
there are no degeneracies other than the Kramers degeneracy.
We will show that 
topological numbers can be defined on this 4D space, 
which we assume to be $S^4$ below\cite{ HatFuk08}.

Owing to the assumption above, 
Kramers doublets are distinguishable on $S^4$, and therefore,
we can examine each Kramers doublet separately.
Let $\Psi$ be a set of eigenfunctions of 
a specific Kramers doublet\cite{HatFuk08}:
\begin{alignat}1
\Psi(x)=\left(\psi(x),-{\cal T}\psi(x)\right) .
\label{KraDou}
\end{alignat}
One finds the time reversal constraint,
\begin{alignat}1
{\cal T}\Psi=\Psi J ,
\label{TimRevCon}
\end{alignat}
where the matrix $J$ is defined by 
\begin{alignat}1
J=\left(\begin{array}{cc}&1\\-1&\end{array}\right)\equiv i\tau^2,
\end{alignat} 
which operates on the space of the Kramers doublet. 
Berry's gauge potential is defined by 
\begin{alignat}1
A&=\Psi^\dagger d\Psi ,
\label{BerGauPot}
\end{alignat}
where $d$ is the exterior derivative with respect to 
$x_a$, $d=dx_a\partial/\partial x_a$.
The time reversal constraint (\ref{TimRevCon}) ensures that $A=-J^{-1}A^TJ$, implying that
$A\in$ sp($1$) algebra. 
In the case of generic $2n$ Kramers multiplet, which may happen if the Hamiltonian has 
some other symmetries, the gauge potential belongs to sp($n$).
Therefore, the first Chern number vanishes since
$\mbox{tr}\,F=0$ holds identically, where $F$ is the field strength 2-form 
defined by $F=dA+A^2$.
Arvon {\it et. al.} \cite{ASSS88} showed that for the time reversal invariant systems, 
the basic topological invariant is the second Chern number,
\begin{alignat}1
c_2=-\frac{1}{8\pi^2}\int_{S^4} \mbox{tr}\,F^2 .
\label{SecCheNum}
\end{alignat}
In the case of the minimum quaternionic 2$\times$2 model mentioned above, 
they derived $c_2=\pm1$ for two Kramers doublets.

The nontrivial Chern number is due to the obstruction of the gauge 
fixing.\cite{Koh85,HGauge04,HatFuk08}
Suppose that we need two patches to span smooth eigenfunctions over $S^4$.
Typical example is such that $S^4$ is divided into two hemispheres which 
are topologically equivalent to the 4D discs, denoted here as $D^4_\pm$.
In the overlap region $D_+^4\cap D_-^4$, 
the two kinds of the wave functions $\Psi_\pm$, defined on $D^4_\pm$, respectively, 
are related with each other through the gauge transformation (the transition function):
\begin{alignat}1
\Psi_+=\Psi_- g .
\label{TraFun}
\end{alignat}
The time reversal constraint (\ref{TimRevCon}) imposes the condition 
$g=J^{-1}g^*J$, which tells that $g\in$Sp($1$). 
The gauge potentials $A_\pm$, defined on the patches $D_\pm^4$, respectively,
are thus related with each other as
\begin{alignat}1
A_+=g^{-1}A_-g+g^{-1}dg .
\label{GauTraA}
\end{alignat} 
As in the case of the first Chern number, the second Chern number 
$c_2$ can be written by the 
transition function $g$ in the following manner:
First, note that 
$\mbox{tr}\, F^2=d\omega_3$, where $\omega_3(A,F)$ is the Chern-Simons 3-form
$\omega_3(A,F)=\mbox{tr}\,\left(AF-A^3/3\right)$.
It is not gauge invariant; 
its transformation law is given by
$\omega_3(A^g,F^g)=\omega_3(A,F)+d\alpha_2-(1/3)\mbox{tr}\,(g^{-1}dg)^3$,
where $A^g$ and $F^g$ denote the gauge transforms of $A$ and $F$ by a generic $g$, 
and $\alpha_2$ is a certain 2-form which vanishes after the integration.
Thus, we find that the second Chern number
$c_2$ defined in eq. (\ref{SecCheNum}) is given by
\begin{alignat}1
c_2=\frac{1}{24\pi^2}\int_{S^3}\mbox{tr}\,\left(g^{-1}dg\right)^3,
\label{SecCheNumTra}
\end{alignat}
where $S^3$ denotes 3-sphere as the boundary of $D_+^4$.
This tells that $c_2$ is given by the winding number of the transition function 
$g$ in eqs. (\ref{TraFun}) and (\ref{GauTraA}) over $S^3$.

So far we have discussed that
the Chern number (\ref{SecCheNum}) would be one of natural topological numbers
for the specification of time reversal invariant systems.\cite{ASSS88}
In more generic cases with $2n$ Kramers multiplet, $g$ belongs to Sp($n$),
and therefore, the Chern number is also given by
a winding number of the transition function $g\in\,$Sp($n$) on $S^3$.
This is ensured by $\pi_3$(Sp($n$))=Z ($n\ge 1$).
All these suggest that time reversal invariant states would be classified by integers
like the IQHE. However, 
it contradicts the $Z_2$ classification of time reversal invariant insulators
by Kane and Mele \cite{KanMel05a}, who assert that 
there are only two phases, trivial insulators and topological insulators 
in two spatial dimensions.
Their argument is quite clear, especially from the point of view of the edge states,
which naturally pick up the topologically nontrivial  property of the bulk:\cite{HatEdge93}
Even number of edge states 
are unstable against time reversal invariant perturbations.
Then, how does the Chern number classification, which corresponds to the bulk picture, 
match the Z$_2$ classification? 
To answer this question, we should note that 
in the above calculations we have used 
the wave function of the Kramers doublet defined in eq. (\ref{KraDou}). 
This equation means that once we choose a wave function $\psi$ for one state, 
we fix that of the other paired state as $-{\cal T}\psi$.
However, we can basically attach an arbitrary phase to the paired state.
This reminds us of
the pfaffian formula proposed by Fu and Kane, \cite{FuKan06} in which the relative 
phase between the Kramers doublet plays an important role.  
Therefore, it is very crucial to take into account of this 
relative phase ambiguity.
To be more precise, this phase is described as follows:
A transformation $\psi\rightarrow e^{i\theta/2}\psi$ induces
\begin{alignat}1
&
\Psi\rightarrow\left(e^{i\theta/2}\psi, -e^{-i\theta/2}({\cal T}\psi)\right)
\equiv\Psi r^{-1}(\theta),
\nonumber\\
&r(\theta)=e^{-i(\theta/2)\tau^3} .
\label{RelPha}
\end{alignat}
For fixed $\theta$, this never affect the topological properties described above.
Indeed, $r(\theta)$ is a gauge transformation belonging to Sp($1$), under which 
the Chern number (\ref{SecCheNum}) is invariant.
However, if the phase $\theta$ is changed adiabatically and 
moved along a loop, 
it reveals a new topological property of the transition function which is 
different from the second Chern number. 
As we will see below, 
such a topological aspect of the transition function is nonperturbative.
Note that $r(0)=1$ and $r(2\pi)=-1$. 
The overall sign of $\pm\Psi$ has nothing to do with the relative 
phase, so that the period of $\theta$ should be $0\le \theta\le2\pi$.
Eq. (\ref{RelPha}) yields the transformation for the transition function,
\begin{alignat}1
g(x)&\rightarrow r^{-1}(\theta)g(x)r(\theta)\equiv g_r(x,\theta) .
\label{RotTraFun}
\end{alignat}
The r.h.s above denotes
the ``rotated'' transition function which now depends on $\theta$ as well as $x$.
Note that the period of $g_r$ with respect to $\theta$ is indeed $2\pi$, 
$g_r(x,0)=g_r(x,2\pi)$. 
Thus, the transition function can be regarded as a function defined on the 4D space
$S^3\times S^1$, where the latter $S^1$ is the loop spanned by $\theta$.
Under the $2\pi$ rotation induced by the adiabatic change of $\theta$, 
the transition function falls into 
two classes, as demonstrated by Witten\cite{Witten83a,Witten83b}. 

Without use of algebraic topology, Witten \cite{Witten83a,Witten83b}
invented a method of calculating
the global anomaly in terms of the perturbative anomaly.
This method is very helpful, also in the present problem, 
to define and to compute the Z$_2$ numbers. 
In what follows, by the use of the technique developed by Witten, 
we first define the Z$_2$ number and next 
show that it is related to the second Chern number $c_2$
such that odd (even) $c_2$ is nontrivial (trivial) element of Z$_2$.
To this end, we embed the Sp($1$)$\simeq$SU($2$) 
transition function $g$ 
into SU($3$) 
such that
\begin{alignat}1
&
\tilde g(x)=\left(\begin{array}{cc}g(x)&\\&1\end{array}\right).
\label{EmbSU3}
\end{alignat}
Likewise, $r(\theta)$ is embedded into SU(3) as 
\begin{alignat}1
\tilde r(\theta)&=e^{-i(\theta/2)(\lambda^3-\sqrt{3}\lambda^8)}
\nonumber\\
&=\mbox{diag}
(1,e^{i\theta},e^{-i\theta}),
\label{TilRBou}
\end{alignat}
where $\lambda^a$ is the standard Gell-Mann matrices.
Here, we have used the fact that the special form 
$\tilde g$ in eq. (\ref{EmbSU3}) is invariant under a particular SU($3$) transformation
$e^{i\alpha\lambda^8}$ with $\lambda^8=\mbox{diag}(1,1,-2)/\sqrt{3}$ for any real $\alpha$.
Although a natural embedding may be only the $\lambda^3=\mbox{diag}(1,-1,0)$
part in eq. (\ref{TilRBou}),
this invariance enables us to introduce another rotation
with respect to $\lambda^8$, as in eq. (\ref{TilRBou}).
Furthermore, we introduce a radial parameter $\rho$ ($0\le\rho\le1$) into $\tilde r$
to enlarge the parameter space $S^3\times S^1$ 
to five dimensional disc-like space $S^3\times D^2$, 
whose boundary ($\rho=1$) is $S^3\times S^1$, i.e, $\partial(S^3\times D^2)=S^3\times S^1$.
At the boundary 
we assume $\tilde r(\theta,1)$ to be eq. (\ref{TilRBou}),
whereas inside $S^3\times D^2$ ($0\le\rho<1$), $\tilde r(\theta,\rho)$ 
is a generic element of SU($3$).
The following is a convenient extrapolation from the boundary to $\rho=0$ proposed by Witten;
\begin{alignat}1
\tilde r(\theta,\rho)=
\left(\begin{array}{ccc}
1&&\\
&\rho e^{i\theta}&\sqrt{1-\rho^2}\\&-\sqrt{1-\rho^2}&\rho e^{-i\theta}\end{array}\right).
\label{RRho}
\end{alignat}
Note that $\tilde r(\theta,\rho=0)$ is independent of $\theta$, 
implying that the space spanned by $(\theta,\rho)$ has indeed a disc geometry.
Thus, the rotated transition function $g_r$ in (\ref{RotTraFun}) 
is embedded into SU($3$) as 
\begin{alignat}1
\tilde g_r(x,\theta,\rho)
=\tilde r^{-1}(\theta,\rho)\tilde g(x)\tilde r(\theta,\rho).
\label{TilG}
\end{alignat}
Now the Z$_2$ number $D$ can be defined by 
\begin{alignat}1
D=\Gamma[\tilde g_r] , 
\label{Z2Inv}
\end{alignat}
where
\begin{alignat}1
\Gamma[\tilde g_r]=
\frac{-i}{240\pi^2}\int_{S^3\times D^2}\mbox{tr}\,\left(\tilde g_r^{-1}d\tilde g_r\right)^5 .
\label{WZW5}
\end{alignat}
To be precise, $D$ in eq. (\ref{Z2Inv}) should be defined mod $2\pi$, as will be clarified momentarily.
In what follows, we study the properties of $D$ thus defined.

First, we note that it is topological, 
since $\Gamma[\tilde g_r]$ is invariant under infinitesimal changes of 
$\tilde g_r$, $\Gamma[\tilde g_r+\delta \tilde g_r]=\Gamma[\tilde g_r]$,
provided that $\tilde g_r$ satisfies the boundary condition (\ref{EmbSU3})
and (\ref{TilRBou}).
We next show that the Z$_2$ number $D$, in the case of the embedding above,
is related to 
the second Chern number. To this end, we note the relation
\begin{alignat}1
\Gamma[gh]=\Gamma[g]+\Gamma[h]+\Delta\Gamma[g,h],
\label{WZIde}
\end{alignat}
for generic SU(3) fields $g$ and $h$. After the integration over $\rho$, 
the last term is given by 
\begin{alignat}1
\Delta\Gamma[g,h]&
=\frac{i}{48\pi^2}\int_{\partial(S^3\times D^2)}
\mbox{tr}\,\Big[
(g^{-1}dg)^3dhh^{-1}
\nonumber\\
&+g^{-1}dg(dhh^{-1})^3+\frac{1}{2}(g^{-1}dgdhh^{-1})^2
\Big].
\label{DelGamIde}
\end{alignat}
Applying the identity (\ref{WZIde}) to eq. (\ref{WZW5}), we obtain
$\Gamma[\tilde g_r]=\Delta\Gamma[\tilde r^{-1},\tilde g\tilde r]
+\Delta\Gamma[\tilde g,\tilde r]$. 
By making use of the fact that $\tilde r(\theta,\rho)$ in (\ref{RRho}) is 
independent of $\theta$ at $\rho=0$, we can further perform the integration over $\theta$,
\begin{alignat}1
\Gamma[\tilde g_r] 
&=\frac{-i}{48\pi^2}\int_{S^3\times S^1}
\mbox{tr}\,d\tilde r\tilde r^{-1}
\left[(d\tilde g\tilde g^{-1})^3+(\tilde g^{-1}d\tilde g)^3\right]
\nonumber\\
&=\frac{1}{24\pi}\int_{S^3}\mbox{tr}\,(g^{-1}dg)^3
\nonumber\\
&=\pi c_2.
\label{Z2VerChe}
\end{alignat}
It thus turns out that 
$D$ defined by eq. (\ref{Z2Inv}) is proportional to the second Chern number. 
In more generic cases with $2n$ Kramers multiplet, 
the above relation (\ref{Z2VerChe}) still holds by embedding Sp($n$) into SU($3n$).

Finally, we study the modulo $2\pi$ property of $D$.  
Basically, it is due to the dependence of the way of embedding in the present formulation.
To show this, it is convenient to introduce another function which denotes 
the relative change of $g_r$ from $g$, 
\begin{alignat}1
G(x,\theta) &=g_r(x,\theta)g^{-1}(x) .
\label{NewMap}
\end{alignat}
This function helps us to see a more precise origin of the Z$_2$ number.
As we have noted, 
$x_a$ and $\theta$ span $S^3$ and $S^1$, respectively, and
therefore, $g_r$ is a mapping from $S^3\times S^1$ to Sp($1$).
On the other hand, the new mapping (\ref{NewMap}) satisfies the relation,
$G(x,0)=G(x,2\pi)=1$. 
This boundary condition as well as
the fact that $\pi_1$(Sp($1$))=0 make it possible to regard  $S^3\times S^1$ as $S^4$,
and thus, the mapping $G$ should be classified by $\pi_4$(Sp($1$))=Z$_2$.
In a similar way above, 
the mapping (\ref{NewMap}) can be embedded into SU($3$) as
\begin{alignat}1
\widetilde G(x,\theta,\rho)=
\tilde g_r(x,\theta,\rho)
\tilde g_r^{-1}(x,0,\rho) ,
\end{alignat}
where $\tilde g_r$ has been defined in eq. (\ref{TilG}).
The boundary condition is 
$\widetilde G(x,0,\rho)=\widetilde G(x,2\pi,\rho)=1$.
Therefore, we can define the Z$_2$ number as 
\begin{alignat}1
D=\Gamma[\widetilde G],\quad \mbox{mod } 2\pi ,
\label{Z2Inv2}
\end{alignat}
where $\Gamma$ is the same as in eq. (\ref{WZW5}) but with the integration domain $D^5$
whose boundary is $S^4$ mentioned-above, $\partial D^5=S^4$.
The modulo $2\pi$ property is now clear: If $D^5$ and $D^{'5}$ are two different discs
but with the same boundary $S^4$, then we have 
\begin{alignat}1
\Gamma_{D^5}[\widetilde G]-\Gamma_{D^{'5}}[\widetilde G]
&=
\frac{-i}{240\pi^2}\int_{S^5}(\widetilde G^{-1}d\widetilde G)^5 .
\end{alignat}
The r.h.s. is manifestly a multiple of $2\pi$, which is ensured by $\pi_5$(SU(3))=Z. 
It is known \cite{Witten83a,Witten83b,EliNai84} 
that when $G$ is a trivial element of $\pi_4$(Sp(1)), 
its embedding $\tilde G$ gives $D=0$, 
whereas in the case with nontrivial $G$ it yields $D=\pi$.
Now, we will show below that two $D$'s defined in eqs. (\ref{Z2Inv}) and (\ref{Z2Inv2})
indeed coincide.
This leads to the conclusion that $g(x)$ on $S^3$ is classified by Z (second Chern number),
but even (odd) elements can 
be continuously deformed into different even (odd) elements through 
the relative phase $\theta$ between the Kramers doublet.
The nontrivial element in $\pi_4$(Sp($1$)), which should describe the nontrivial topological 
states with time reversal symmetry, is thus given by $g(x)$ with odd second Chern numbers. 
We believe that this is the origin of Z$_2$ characteristics of the topological number
specifying the phases of time reversal invariant systems.

At first sight, the two definitions (\ref{Z2Inv}) and (\ref{Z2Inv2}) seem different, but
straightforward calculations lead us to 
$\Gamma[\widetilde G]=\Gamma[\tilde g_r]$.
To show this, we note that eq. (\ref{WZIde}) yields
\begin{alignat}1
\Gamma[\widetilde G]
&=\Gamma[\tilde g_r]+\Delta\Gamma , 
\end{alignat}
where the second term is essentially given by eq. (\ref{DelGamIde}) with
$g\rightarrow \tilde g_r(x,\theta,\rho)$ and $h\rightarrow g_r^{-1}(x,0,\rho)$,
\begin{alignat}1
\Delta\Gamma& 
=
\frac{-i}{48\pi^2}
\int_{S^3\times S^1}\mbox{tr}\,
\Big[
(\tilde g_r^{-1}d\tilde g_r)^3\tilde g^{-1} d\tilde g
\nonumber\\
&+
\tilde g_r^{-1}d\tilde g_r(\tilde g^{-1} d\tilde g)^3
-\frac{1}{2}(\tilde g_r^{-1}d\tilde g_r\tilde g^{-1} d\tilde g)^2
\Big] .
\label{DelGam}
\end{alignat}
In what follows, we show that this term vanishes, $\Delta\Gamma=0$, and 
hence, $\Gamma[\widetilde G]=\Gamma[\tilde g_r]$ holds. 
To this end, we first note that the 1-form $\tilde g^{-1}d\tilde g$ 
is written as
\begin{alignat}1
\tilde v&\equiv
\tilde g^{-1}d\tilde g 
=\left(
\begin{array}{cc}
v&\\ & 0
\end{array}
\label{TilV}
\right) ,
\end{alignat}
where $v$ is sp(1)-valued 1-form $v\equiv g^{-1}dg$. 
Then, one finds that the rotated 1-form can be decomposed into two parts:
\begin{alignat}1
\tilde g_r^{-1}d\tilde g_r=\tilde u+\tilde v_r .
\end{alignat}
The former is the 1-form $\tilde u$ including only $d\theta$, while the latter is 
rotated $\tilde v$ with $dx$ only,
\begin{alignat}1
\tilde u&\equiv\tilde r^{-1}\tilde g^{-1}\tilde rd\tilde r^{-1}\tilde g\tilde r+
\tilde r^{-1}d\tilde r
\nonumber\\
&=\frac{i}{2}(\tilde r^{-1}\tilde g^{-1}\lambda^3\tilde g\tilde r-\lambda_3)d\theta
\nonumber\\
&=\left(
\begin{array}{cc}
ud\theta
&\\ & 0
\end{array}
\right) ,
\nonumber\\
\tilde v_r&=\tilde r^{-1}\tilde v\tilde r
=\left(
\begin{array}{cc}
v_r &\\ & 0
\end{array}
\right) ,
\label{AlpTilV}
\end{alignat}
where $u=\frac{i}{2}r^{-1}(g^{-1}\tau^3g-\tau^3)r=u^a\tau^a $ 
and $v_r=r^{-1}vr$.
Now eqs. (\ref{TilV}) and (\ref{AlpTilV}) tell that the fields in 
the integrand in eq. (\ref{DelGam}) can be written only by sp(1)-valued fields,
\begin{alignat}1
\Delta\Gamma
=&
\frac{-i}{48\pi^2}
\int_0^{2\pi} d\theta\int_{S^3}\mbox{tr}\,u
\left(
v_r^2v
+vv_r^2
+v^3-v_rvv_r-vv_rv
\right).
\label{DelGam2}
\end{alignat}
The first three terms in the integrand vanish under the trace
because sp(1) is pseudo-real.
The last two terms seem to be complicated, because
$\theta$ and $x$ are coupled together. 
Nevertheless, $\theta$ is basically decoupled from $x$, and therefore,
we can carry out the integration over $\theta$. 
To see this, let us denote $v=v^a\tau^a$, and calculate,
for example, the following term;
\begin{alignat}1
v_rvv_r&=r^{-1}vrvr^{-1}vr
\nonumber\\
&=v^1v^2v^3\epsilon_{abc}r^{-1}\tau^a r\tau^br^{-1}\tau^cr
\nonumber\\
&=v^1v^2v^3\tau^1\tau^2\tau^3 \,2(r^2+r^{-2}+1)
\nonumber\\
&=\frac{2}{3!}v^3(r^2+r^{-2}+1) .
\end{alignat}
The same formula holds for the other term in eq. (\ref{DelGam2}), and we finally arrive at
\begin{alignat}1
\Delta&\Gamma
=\frac{i}{48\pi^2}\frac{4}{3!}
\int_0^{2\pi} d\theta\int_{S^3}\mbox{tr}\,
u\left[
v^3(r^2+r^{-2}+1)
\right] .
\end{alignat}
Since $v^3$ is proportional to the identity matrix, we find that 
the terms with $r^2$ and $r^{-2}$ vanish by integration over $\theta$.
The same is true for the last term because of
the identity $\mbox{tr}\,\tau^av^3=0$. Thus, we conclude that $\Delta\Gamma=0$,
and therefore, $\Gamma[\widetilde G]=\Gamma[\tilde g_r]$.
It follows from eq. (\ref{Z2VerChe}) that for trivial and 
nontrivial mapping $G=r^{-1}grg^{-1}$, $g$ has even and odd 
second Chern number $c_2$, respectively.

In summary, we have shown that the second Chern number is relevant
topological number for the time reversal invariant systems with 
${\cal T}^2=-1$. However, the Kramers doublet has an ambiguity in relative 
phase between the wave functions of the doublet. Through the adiabatic change of this phase,
the wave function with a certain Chern number can be continuously
deformed into other wavefunctions with different Chern numbers.
What is important is that in this deformation the evenness and the oddness 
of the Chern numbers are conserved. 
We believe that this is the origin of the Z$_2$ property of the topological 
number for the time reversal invariant systems. 
Mathematically, it can be described by the fourth homotopy class of the 
transition function with the relative phase degree of freedom, $\pi_4$(Sp($n$))=Z$_2$.

One of Authors (T. Fukui) would like to thank H. Oshima for fruitful discussions. 
This work was supported in part 
by Grant-in-Aid for Scientific Research  
(Grant No. 20340098) from JSPS.
The work by YH was also supported in part by Grants-in-Aid for Scientific Research,
No. 20654034 from JSPS and
No. 220029004 (physics of new quantum phases in super clean materials)
and 20046002 (Novel States of Matter Induced by Frustration) on Priority Areas from MEXT.

\end{document}